\documentclass[aps,pra,twocolumn,showpacs,floatfix,superscriptaddress,nofootinbib]{revtex4}
\usepackage{graphicx}
\usepackage{hyperref}
\usepackage{amsmath}
\usepackage{amssymb}
\newcommand{\ket}[1]{|#1\rangle}
\newcommand{\bra}[1]{\langle#1|}

\def\>{\rangle}
\def\<{\langle}

\begin{document}

\title{How good must single photon sources and detectors be for efficient linear optical quantum computation?}

\author{Michael Varnava}
\address{QOLS, Blackett Laboratory, Imperial College London, Prince Consort Road, London SW7 2BW,
UK}

\author{Daniel E. Browne}
\address{Departments of Materials and Physics, University of Oxford, Parks Road, Oxford, OX1 3PH, UK}

\author{Terry Rudolph}
\address{QOLS, Blackett  Laboratory, Imperial College London, Prince Consort Road, London SW7 2BW,
UK}
\address{Institute for Mathematical Sciences, Imperial College London, 53 Exhibition Road,
London SW7 2BW, UK}

\begin{abstract}
We present a scheme for linear optical quantum computation (LOQC)
which is highly robust to imperfect single photon sources and
inefficient detectors. In particular we show that if the product of
the detector efficiency with the source efficiency is greater than 2/3,
then efficient LOQC is possible. This threshold is many orders of
magnitude more relaxed than those which could be inferred by
application of standard results in fault tolerance. The result is
achieved within the cluster state paradigm for quantum computation.
\end{abstract}


\maketitle

Photons are a promising candidate for quantum information processing
\cite{KokLOQCreview}. They are relatively immune to decoherence, allow a high experimental repetition rate and  high precision single-qubit operations. There are, however, a number of barriers to
building a scalable photonic quantum computer. Entangling two-qubit
operations are challenging to implement, requiring either a highly
non-linear material \cite{imoto} (with strength many orders of magnitude higher
than any known materials \cite{barrettweaknon}) or by employing linear optical elements and
``measurement-induced non-linearities'' \cite{klm}. In this latter alternative, combining a
photo-detection device with linear optical elements such as
beam-splitters, phase shifters and polarizing beam splitters allows
for non-deterministic two-qubit operations which enable efficient linear optical quantum computing (LOQC).

Recently there have been significant experimental advances in LOQC \cite{LOQCexp,KokLOQCreview} but three major experimental issues remain - imperfect sources,
inefficient detectors, and poor quantum memory; our results have
significant implications for all three issues. In particular, our main result is to show that efficient LOQC is possible providing  the detector efficiency $\eta_D$ and the single photon source efficiency $\eta_S$ (defined precisely below) satisfy:
\[\eta_S\, \eta_D>2/3.\]
This threshold is greatly relaxed from what can be inferred from standard fault tolerance results (which imply necessary efficiencies several orders of magnitude higher than our result) - the key difference is the ability to compensate for detected errors using correlations which arise naturally in measurement based quantum computation. In a subsequent publication \cite{inprep} we will show in some detail how to construct good quantum memory using the techniques we introduce here.

Recently, there have been a number of promising proposals for optical quantum computation under a degree of experimental error \cite{dawsonhasel,ralph}. Here, we show a significant reduction in the threshold required for scalable linear optical quantum computation with inefficient sources and detectors.
Similarly to in other work which has addressed methods of reducing the effects of such errors by linear optical means \cite{ralph},  we will be assuming that the
linear optical elements of the computation are ideal, except
possibly for some amount of absorptive loss.

Certain photon sources are of the form that if any photons are
present, then at most one photon is present (see e.g.
\cite{fransonetal}). For such sources the desired mode is precisely known, and it is
hoped the emitted photon's wavepacket has significant overlap with
this mode. Because the transformation between mode operators is
linear, there exists a linear optical filter which can be used to
``project out'' (absorb) the undesired part of the photonic
wavefunction. We call the probability of the photon surviving
this filtration $\eta_S$, the efficiency of the photon source. More
precisely, the mixed state of the source is taken to be
$\rho=\eta_S|1\>\<1|+(1-\eta_S)|0\>\<0|,$with $|1\>$ and $|0\>$ the single photon and vacuum Fock states respectively.

The single photon detectors we consider are number resolving and
inefficient (although the vast majority of the photo-detectors in our protocol can be non-number resolving). There are two primary technologies for such detectors under active investigation:
visible light photon counters \cite{VLPC,loopy}. and
transition edge superconducting sensors \cite{higheffdet}.
Inefficiencies in these detectors are well modeled as an initially
perfect detector, in front of which is placed a regular beamsplitter
of transmission probability $\eta_D$; photons reflecting off the
beamsplitter are presumed lost \cite{KokLOQCreview}. Thus there is a
probability that two input photons register as only one photon - a
particularly problematic scenario for LOQC gates which are generally
conditioned on detection of one and only one photon.

The method we propose to achieve the main result is based on the measurement-based ``one-way quantum
computation''  \cite{clusterqc1}, in particular the
loss-tolerant variant introduced by us in \cite{varnava}.
In one-way quantum computation, single-qubit measurements on a
cluster state of suitable layout and size suffice to implement any
quantum computation. In \cite{varnava} we showed that if cluster
states are encoded using branched tree-structures to represent each
cluster qubit, efficient computation is possible with an overall
loss rate of up to 50\%.

The results of \cite{varnava} apply to the case where a large cluster state has been constructed, and independent (uncorrelated) qubit loss now affects the qubits. However, in the LOQC scenario under consideration we are beginning with imperfect sources of qubits (polarized photons), and attempting to efficiently grow large clusters by implementing noisy gates upon them. It is crucial that the method is not only efficient, but also produces cluster states of an independently degraded (ID) form. That is, the states must be equivalent to an initially ideal cluster state of which each qubit has passed through an independent (uncorrelated) loss channel.  This is because the loss tolerant protocol in \cite{varnava} is much less robust to correlated errors.

Our description of a strategy to achieve the main result proceeds as follows. First we present a method for creating 3 photon ID-GHZ states from an initial resource of 6 single photons. The effective loss rate of this state is a function of the efficiencies of the sources/detectors used in its creation, and this is the only step in which number-resolving detectors are required. Next, we discuss the manner in which Type-II fusion is inherently loss tolerant - when successfully applied to ID cluster states it generates an ID state with the same effective loss rate as the input states. After a brief discussion of how tree cluster states achieve their loss tolerant properties we present a method to efficiently grow such states given the initial 3 photon ID GHZ states. We then describe a method for joining such tree clusters so as to produce a state capable of achieving universal quantum computation.

\begin{figure}
\begin{center}\includegraphics[width=8cm]{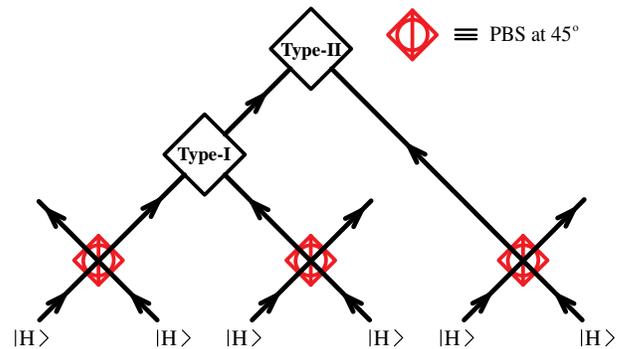}\end{center}
\caption{ \label{makeghz} A simple procedure for projecting 6 single
photons into a three photon GHZ state. Success occurs when three of
the detectors detect a single photon.}
\end{figure}

\noindent\textit{Creating initial GHZ states:} A scheme for creating an initial resource of GHZ states is given in Fig.~\ref{makeghz}.
Whenever three photons are detected at any three of the detectors,
the remaining three photons are projected into a state locally
equivalent to the three photon GHZ state. Heuristically the
interferometer in this figure can be understood as follows: the
three initial pairs of photons go through a rotated PBS, so that
the one-photon per mode part of their state is a $|\phi^+\>^{\otimes
3}$ state. This will be post-selected and simultaneously fused by the
remaining operations. The first pair of Bell states is Type-I fused,
then one member of the second Bell pair undergoes a Type-II
operation with one member of the third Bell pair.

Recall that we consider sources which are of the form that if any
photons are present, then at most one photon is present. Let the
efficiency of the sources (as defined in the introduction) be
$\eta_S$, and that of the detectors be $\eta_D$. In the particular
scheme of Fig.~\ref{makeghz}, if the gate is successful then
somewhat tedious calculations (c.f~APPENDIX~A) reveal that the state
of the remaining three photons are collapsed to an ID GHZ state
which has a loss rate $\epsilon=1-\frac{\eta_S}{2-\eta_D \eta_S}$.
In the rest of the paper we show that loss-tolerant cluster states
with an ID loss rate of $\epsilon$ can be constructed efficiently.
Since, once we take the inefficiencies of single-qubit measurements
into account, the  protocol \cite{varnava} has a loss threshold of
$(1-\epsilon) \eta_D>1/2$ this implies our main result.

The reason we require GHZ states initially (as opposed to Bell states, which sufficed in \cite{tezdan}) is that all subsequent steps of the protocol are going to use Type-II fusion, which destroys two photons on each application (Type-I fusion unfortunately does not have the same natural loss tolerance properties). Fusing 3-photon GHZ states allows for the creation of four-photon (and larger) GHZ states by postselection on successful fusions in a straightforward manner, and it is then clear that if there are no lossy failures, arbitrary cluster states can be built using Type-II fusion alone. If losses during the construction of the state are considered, however, growing suitable clusters efficiently is rather more complex.

\noindent\emph{Loss tolerance of Type-II fusion:}
Recall the Type-II fusion gate is essentially a polarizing beamsplitter (PBS) oriented at $45^\circ$ \cite{braunstein,fransonparity,tezdan}, which implements a destructive  Bell state projection when it succeeds. Success occurs when one (and only one) photon is detected at two independent detectors in the output modes of the gate. We classify as failure all other detection outcomes. Such failures can arise when two photons emerge in a single mode, or when loss or detector inefficiency causes only one photon to be detected. 

In the ideal case, the success probability of the Type-II fusion is
50\%. Provided there is at most one photon in each mode, this gate
is robust to loss errors and detector inefficiencies, because when two detectors click the desired case of one photon being present in each input mode \emph{must} have been the case. The effect of
the loss errors (imperfect sources, detectors or absorptive components) will be to reduce the success probability. Significantly, however, a successful Type-II preserves the
ID property of the input state. More specifically, successfully fusing an
$n$-photon ID state with an $m$-photon ID state, both with loss rate $\epsilon$, leaves an $n+m-2$ photon state which remains ID with the \emph{same}
loss rate $\epsilon$. In other words, the Type-II gate does not introduce
dangerous correlated loss errors into the state, nor does it increase the loss rate, even though it is implemented with inefficient detectors.

\noindent\emph{Tree-clusters: a resource for loss-tolerant computation}:
We now briefly review the manner in which tree clusters achieve their loss tolerant properties.
These states are fully specified by
their branching parameters $b_{i}$ which give the number of qubits
in the $(i+1)^{th}$ level a qubit in the $i^{th}$ level branches
out to. When such trees are used to encode qubits in a cluster state used for a computation (see \cite{varnava} for details), then a plethora of
alternative measurement patterns  become available
for implementing a specific encoded single qubit measurement in some arbitrary basis. This allows a loss tolerant strategy which, by increasing the number of tree qubits, can boost the effective success probability for the measurement arbitrarily close to one, provided each individual measurement has an individual likelyhood to succeed of at least 1/2.

The basis of this loss tolerance is ``counter-factual'' error correction. At instances where
 a specific qubit is lost,  the special
quantum correlations present on the tree cluster states allow ``indirect measurements''. In other words the  outcome of a measurement which could have occured, had the qubit not been lost, can be inferred by measurements on other surviving qubits.
 One can then proceed with an
alternative measurement pattern which is still suitable for
implementing the original logical operation.

\noindent\emph{Building tree-clusters in the presence of loss:}
Type-II fusing one photon from each of two 3-photon GHZ
states yields, when successful, a 4-photon GHZ state, and we will take these to be our basic resources for tree cluster creation.
%

\begin{figure}[ht]
    \includegraphics[width=7.5cm]{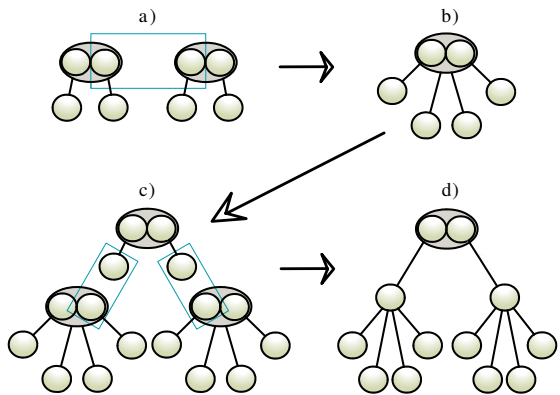}
    \caption{\label{strategy} A strategy for creating tree-clusters via Type-II fusion (denoted by a blue-box).}
\end{figure}

By applying local Hadamard gates to
two of the qubits in a 4-qubit GHZ state, one creates a three-qubit
tree cluster state, where the central qubit is redundantly encoded in the
two-qubit basis $\ket{HH},\ket{VV}$. We will term this basic resource a ``2-tree''. Figure \ref{strategy}(a) and (b) indicates how two 2-trees
can be Type-II fused together into a
4-tree, maintaining the redundant encoding at the top of the tree.

Given a supply/resource of 2-trees, we wish to show that one can
efficiently generate arbitrary tree cluster states. This is achieved by building a
tree with branching parameters \{$b_{0},b_{1},...b_{m}$\} from
bottom to top as follows.

First we use Type-II gates and combine 2-trees to form 4-trees.
Then we repeat
the process by fusing 4-trees to create 8-trees, and so on
until a sufficient resource of $b_{m}$-trees is created. Any given Type-II fusion succeeds with probability $p_{II}\equiv(1-\epsilon)^2\eta_D^2/2$.  This is because $(1-\epsilon)^2$ is the probability of both photons being present in their appropriate cluster (which is given above for one specific proposal for producing GHZ states), $\eta_D^2$ is the probability both detectors fire appropriately, and $1/2$ is the intrinsic optimal success probability of a Type-II gate even if both photons are present and the detectors are perfect.
Thus, for any
integer number $l$, the expected cost for creating one
$2^{l}$-tree from two $2^{l-1}$-trees is $ 2/p_{II}$
$2^{l-1}$-trees.
 The expected cost for creating a
$2^{l}$-tree from 2-trees is then $
(2/p_{II})^{l-1}$ 2-trees. Further, it can be readily shown that in
order to create a $b_{m}$-tree such that $2^{l-1}\leq b_{m} \leq
2^{l}$ ,then on average the number of 2-trees required is
$\leq(2/p_{II})^{log_{2}(b_{m})}= $poly($b_{m}$). (Note that in the event of a failure - regardless of the type - we simply discard all qubits involved. While clearly not optimal from a resource perspective, this suffices to show the desired scaling to attain a proper fault tolerant threshold.)

Next we add a higher level of qubits by first joining a pair
of the cluster states created in the previous step with a single
2-tree using two Type-II gates as shown on Figure \ref{strategy}
 (c). Subject to both gates succeeding, the resulting state is
be the one shown on figure~\ref{strategy}(d). This now is a tree
cluster state consisting of a redundantly encoded qubit at the top
that is branching out to 2 qubits in the next lower level, each of
which branch out to $b_{m}$ qubits in the last level. Since we
require two Type-II gates to succeed in this step, the
expected number of 2-trees consumed in order to create a single
such state is $\leq 2 p_{II}^{-2} \textrm{poly}(b_{m}$).

By creating a sufficient supply of these new cluster states we can now
increase the branching parameter on the top level from 2 to
$b_{m-1}$ by combining these tree clusters together, much as we combined the initial 2-trees into $2^l$ trees. That is, successfully
Type-II fusing together a photon from the redundantly encoded qubit
from  2 of these trees creates new tree clusters where the
top level branching would then be equal to 4; fusing those
together increases the branching to 8 etc. Repeating the
process can increase the branching value to $b_{m-1}$. This increases the resource overhead in the
number of 2-trees required to
$\leq(2/p_{II})^{log_{2}(b_{m-1})}2 p_{II}^{-2}$poly($b_{m}$).
The extra added  level with branching
parameter $b_{m-1}$ incurs an increasing factor of
$2 p_{II}^{-2}$poly($b_{m-1}$) in the 2-trees overhead.

Iterating the process in order to add all the levels  suggests
that in order to create one tree cluster state with the full
branching parameter profile: \{$b_{0},b_{1},...b_{m}$\} (as required
in \cite{varnava}) then the expected number of 2-trees required
satisfies
\begin{equation}\label{no2trees}
\langle N_\textrm{tree}\rangle \leq \left(\frac{2}{p_{II}^2}\right)^{\!\! m}\prod\limits_{i=0}^{m}\textrm{poly}(b_{i}).
\end{equation}
The overall conclusion is that  the expected number of qubits consumed in order to build a tree containing $Q$ qubits is polynomial in $Q$, since $m\leq log_{2}(Q)$.

%
%

\begin{figure}[ht]
\begin{center}
    \includegraphics[width=8cm]{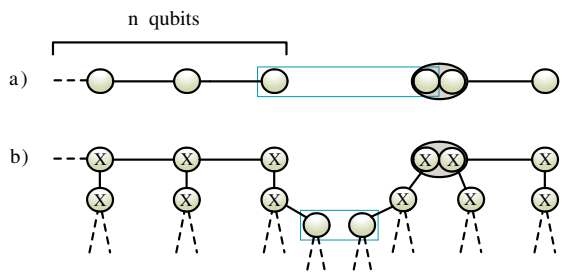}
    \end{center}
    \label{treestocluster}
    \caption{In figure (a), the Type-II fusion gate  (denoted by a blue-box) between to the end qubit of an $n$-qubit cluster and a redundantly encoded qubit from a three qubit cluster state will ``fuse'' the states together, generating, in this example an $n+1$-qubit cluster. Figure (b) presents the loss encoded version of this protocol. Note that the qubits marked with an $x$  have already been measured in the $\sigma_x$-basis. They have been left in the diagram for illustrative purposes. The graphical representation of the true loss-tolerant state after the $x$-measurements would be unhelpfully complicated.
        }
\end{figure}

\noindent\emph{From trees to a loss tolerant cluster state}: The final step of our protocol is to combine the generated trees
into an encoded cluster state. This is achieved using a strategy similar, at the level of encoded qubits, to those strategies proposed in \cite{tezdan}. We start with an $n$-qubit linear cluster and a $2-$qubit cluster state, of which  qubit is redundantly encoded as illustrated in Figure \ref{treestocluster} (this is locally equivalent to a 3-qubit GHZ state). Successful application of a Type-II fusion gate between  qubits creates an $n$+1 linear cluster. A similar approach can be employed to create loss-encoded linear clusters. The Type-II fusion on the encoded qubits proceeds in an analogous manner to the loss-tolerant single qubit measurements in \cite{varnava}. The fusion measurement is applied to a pair of ``first row'' qubits, as illustrated in Figure \ref{treestocluster}, and remaining tree qubits are measured out in accordance to the loss tolerant strategy \cite{varnava}. Loss errors in the Type-II fusion are dealt with in the same way as before. When the Type-II fusion gate fails it can be treated as a loss error. Note that this means that at the level of encoded qubits the failure probability of the  fusion can be made arbitrarily small.

Thus an $n$ encoded qubit linear cluster can be built at a cost of $n$ tree-encoded three-GHZ states. Following the methods above, these states can themselves be built (for example) by post-selected fusion of three tree clusters and a 4-photon linear cluster state at a cost of $(3/p_{II}^3+3\langle N_\textrm{tree}\rangle)/p_{II}^3$ 2-trees.

\noindent\emph{Conclusion:} Our primary purpose in this paper has
been to show that an extremely relaxed error threshold exists for
some of the primary error mechanisms expected to be crucial to the
eventual viability of linear optical quantum computation.


It is possible that our threshold trade-off can be improved - fall
that is required is a better linear optical scheme for producing
three photon ID GHZ states than the one we have outlined, which we
have made no attempt to optimize.

There are clearly many ways in which the resource consumption of our scheme can be lowered. In the methods above we have made no attempts to recycle states after gate ``failures'' and to do so would greatly reduce the overhead, especially when efficiencies are significantly above the threshold.
Although the experimental requirements of the current scheme seem forbidding, we are confident that experiments demonstrating the principles of elements of this schemes are within current laboratory feasibility.
The development of loss tolerant protocols with a more compact resource count remains an important and challenging area for future research.

\section*{ACKKOWLEDGEMENTS}
We acknowledge useful conversations with Chris Dawson and Jens Eisert. This research was supported by  DTO-funded U.S. Army Research Office Contract No.
W911NF-05-0397, Merton College, Oxford and the Engineering and
Physical Sciences Research Council (EPSRC) and the EPSRC's QIPIRC.

\begin{appendix}

\section{Creating ID-GHZ states}

\begin{figure}[h]
\begin{center}\includegraphics[width=8.5cm]{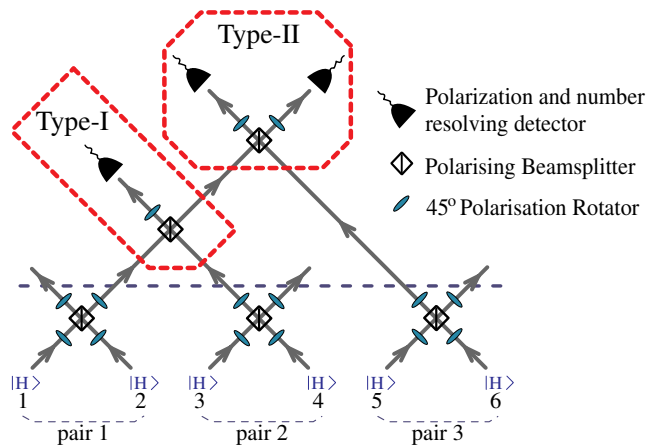}\end{center}
\caption{ \label{circuit} Circuit for producing a 3-photon ID-GHZ
state out of 6 unentangled photons at the input.}
\end{figure}

Consider Fig.~\ref{circuit} where we depict a gate that operates on
6 horizontally polarized photons, one at each input of the six
spatial modes.

The objective will be to show that given that this particular gate
can produce a three photon ID-GHZ state at the output which takes
the form:
\begin{eqnarray}\label{ghz}
&&ID\_GHZ= (1-f)^3 \ket{GHZ}\bra{GHZ}\nonumber \\
&&+\tfrac{(1-f)^2f}{2}(\ket{H_1H_2}\bra{H_1H_2}+\ket{V_1V_2}\bra{V_1V_2}\nonumber \\
&&~~~~~~~~~~~~~~\ldots+\ket{V_2V_3}\bra{V_2V_3}) \nonumber\\
&&+\tfrac{(1-f)f^2}{2}\left(\ket{H_1}\bra{H_1}+\ket{V_1}\bra{V_1}\ldots+\ket{V_3}\bra{V_3}\right) \nonumber\\
&&+f^3\ket{vac}\bra{vac}\textrm{,}
\end{eqnarray}
where $\ket{GHZ}=\frac{\ket{HHH}+\ket{VVV}}{\sqrt{2}}$.

Eqn.~\ref{ghz} gives the mixed state which would result if a
3-photon GHZ state is initiated perfectly, but subsequently each of
the photons is lost at a probability rate $f$. To show that the gate
in Fig.~\ref{circuit} produces such a state, we will first assume
that losses can only occur at the input, due to imperfect sources
which have efficiency $\eta_S<1$. We will assume that the
polarization-sensitive, number-resolving detectors are perfect and
show that the gate can produce a 3 photon ID-GHZ. Later on we will
show that this scenario is exactly equivalent to a scenario where we
have both imperfect sources and imperfect detectors.

The circuit shown Fig.~\ref{circuit} post-selects only the outcomes
where there is one and only one photon detected at each of the
separate 3 detectors. We will now attempt to construct the mixed
state that is received at the output when the detection pattern at
the detectors corresponds to such correct detection outcomes. The
convention that we will be using throughout for the labeling of the
spatial modes in the circuit is that they continue in straight
lines. In other words a spatial mode is transmitted through a
polarizing beamsplitter (PBS).

We would investigate 4 distinct cases, where in each case a
different number of photons is lost. These are the cases where 0, 1,
2 or 3 photons are lost. Obviously if 4 or more photons are lost it
is impossible to post-select a correct detection pattern. Since we
have lossy components, we expect the final state to be a mixed state
of the form:

\begin{equation}\label{mix}
\rho_{out}=P_3\ket{3}\bra{3}+P_2\ket{2}\bra{2}+P_1\ket{1}\bra{1}+P_0\ket{0}\bra{0}\textrm{,}
\end{equation}

where $P_{i}$ is the re-normalized probability of having an
$i$~-~photon state in the final mixture given by the
$\ket{i}\bra{i}$ density operator. Ultimately we would like to show
that $\rho_{out}$ takes the form of Eqn.~(\ref{ghz}) as this would
be proof that the circuit does indeed produce an ID-GHZ state at the
output.

\subsection*{Case I: 0 photons lost} First we will look at the state
as soon as all 6 photons pass through the 3 PBS's rotated at
$45^\circ$. From now on we will use the notation PBS45$^\circ_{ij}$
to refer to a PBS placed on spatial modes labeled $i$ and $j$. The
action of PBS45$^\circ_{ij}$ acting on a pair of photons in state
$\ket{H_iH_j}$ is:
\begin{eqnarray}\label{T2}
\ket{H_iH_j}&\rightarrow& \frac{1}{2\sqrt{2}}\left(
\ket{H_{1}^{2}0_2}-\ket{V_{1}^{2}0_2}+\ket{0_1H_{2}^{2}}-\ket{0_1V_{2}^{2}}
\right) \nonumber \\
&&+~\frac{1}{2} \left( \ket{H_iH_j} +\ket{V_iV_j}\right)\textrm{.}
\end{eqnarray}

Clearly the terms which lead to 2 photons in the same spatial mode
cannot contribute to a successful detection pattern since in such
instances the photons have the same polarization and are destined to
end up at the same detector, triggering an erroneous detection
outcome. Thus the only terms that survive to (possibly) lead to a
correct detection pattern are the ones which have just one photon at
each separate spatial mode.

Thus after the action of the three PBS45$^\circ$'s (c.f dotted line
on Fig.~\ref{circuit}) the state which possibly yields a correct
detection pattern is given by:
\begin{equation}
\frac{\left(\ket{H_1H_2} +\ket{V_1V_2}\right)\left(\ket{H_3H_4}
+\ket{V_3V_4}\right)\left(\ket{H_5H_6}
+\ket{V_5V_6}\right)}{8}\textrm{.}
\end{equation}

At this point we need to consider how the state evolves after the
action of the two PBS's mixing modes 1 and 4, and modes 1 and 6
respectively. From now on we will use the notation PBS$_{ij}$ to
refer to a PBS mixing modes $i$ and $j$. The convention we will use
for any PBS in the circuit is that it transmits horizontally
polarized photons and reflects vertically polarized photons. This
can be summarized as follows:

\begin{eqnarray}
PBS_{ij}:& H_i \rightarrow H_i\textrm{,}& \nonumber \\
&H_j \rightarrow H_j\textrm{,}& \nonumber \\
&V_i \rightarrow V_j\textrm{,}& \nonumber \\
&V_j \rightarrow V_i\textrm{.}&
\end{eqnarray}

After the action of PBS$_{14}$ the new state takes the form:
\begin{eqnarray}
&&\frac{1}{8}\left(\ket{H_1H_2H_3H_4}+\ket{V_4V_2H_3H_4}+\ket{H_1H_2V_3V_1}\right.\nonumber \\%
&&~\left.+\ket{V_1V_2V_3V_4}\right)\left(\ket{H_5H_6}
+\ket{V_5V_6}\right)\textrm{.}
\end{eqnarray}

Clearly the term involving $\ket{V_4V_2H_3H_4}$ cannot contribute to
a correct detection outcome as it leads with certainty to an outcome
involving 2 photon detections at the detector in spatial mode 4.
Similarly neither the term involving $\ket{H_1H_2V_3V_1}$ can lead
to a correct detection outcome as it leads with certainty to an
outcome with no detections at the detector in mode 4. Since we only
post select the results involving one photon being detected at this
detector then the part of the state which may lead to such an
outcome is:
\begin{eqnarray}
\frac{1}{8}\left(\right.\ket{H_1H_2H_3H_4H_5H_6}+\ket{V_1V_2V_3V_4H_5H_6}\nonumber
\\
\ket{H_1H_2H_3H_4V_5V_6}+\ket{V_1V_2V_3V_4V_5V_6} \left.\right)
\textrm{.}
\end{eqnarray}

After the action of the PBS$_{16}$ this becomes:
\begin{eqnarray}\label{beforerotators}
\frac{1}{8}\left(\right.\ket{H_1H_2H_3H_4H_5H_6}&+&\ket{V_6V_2V_3V_4H_5H_6}\nonumber
\\
+~\ket{H_1H_2H_3H_4V_5V_1}&+&\ket{V_1V_2V_3V_4V_5V_6} \left.\right)
\textrm{.}
\end{eqnarray}
Again we can see that the terms involving $\ket{V_6V_2V_3V_4H_5H_6}$
and $\ket{H_1H_2H_3H_4V_5V_1}$ cannot contribute to a correct
detection outcome since the first term leads with certainty to 2
photons be detected at the detector in spatial mode 6 whereas the
second leads to 2 photons to be detected at the detector in mode 1.
Thus the only terms from this state which could lead to a correct
detection pattern are:
\begin{equation}
\frac{1}{8}\left(\right.\ket{H_1H_2H_3H_4H_5H_6}+\ket{V_1V_2V_3V_4V_5V_6}
\left.\right) \textrm{.}
\end{equation}
By applying the three 45$^\circ$-polarization rotators found prior
to the three detectors at modes 1, 4 and 6 we find that the only
term involving $\ket{H_1H_4H_6}$ that survives is:
\begin{eqnarray}\label{HHH0}
&\frac{\ket{H_1H_4H_6}}{16\sqrt{2}}&\left(\ket{H_2H_3H_5}+\ket{V_2V_3V_5}
\right)\nonumber \\
=&\frac{\ket{H_1H_4H_6}}{16}&\ket{GHZ}_{235} \textrm{.}
\end{eqnarray}
This is one possible correct detection pattern out of a total of 8
possible correct detection patterns (e.g. $\ket{H_1V_4H_6}$,
$\ket{H_1V_4V_6}$ etc. will work as well). Eqn.~(\ref{HHH0})
suggests that the probability with which the initial 6 photon state
ends up with $\ket{GHZ}_{235}$ at the output with all three
detectors firing with one horizontally polarized photon is 1/256.
Note that one can easily show that the remaining possible correct
detection outcomes occur with the same probability of 1/256 and they
give rise to states that are locally equivalent to
$\ket{GHZ}_{235}$. Thus assuming that the local corrections are
applied the overall success probability for the gate to produce the
GHZ state is 1/32.

Since the probability of having 6 photons at the input is given by
$\eta_S^6$ suggests that $\ket{3}\bra{3}$ in Eqn.~(\ref{mix}) is
given by:
\begin{equation}
\ket{3}\bra{3}=\frac{\eta_S^6}{32}\ket{GHZ}_{235}\bra{GHZ}_{235}\textrm{.}
\end{equation}

\subsection*{Case II: 1 photon lost} We will now look at the case
where one photon is lost, and initially also assume that this photon
comes from the pair of photons initially found in input modes 1 or
2. Let us define the two variables, $p_1$ and $p_2$ such that the
value of any given $p_i$ is 0 if the photon in input mode $i$ is not
present and 1 if indeed it is present.

In other words if we want to check the case where one and only one
photon is lost from the pair of input modes 1 and 2. The variables
$p_1$ and $p_2$ can be described by the equations:
\begin{eqnarray}\label{onelost}
p_1+p_2&=&1\textrm{,} \nonumber \\
p_1p_2&=&0\textrm{.}
\end{eqnarray}
The equations above make sure that one photon is lost from the pair
of modes 1 and 2. Note there are 2 possible solutions.

Initially the 5 qubit state is taken to be one of the two possible
states given by:
\begin{equation}\label{p1p2}
(p_1\ket{H_1}+p_2\ket{H_2})\ket{H_3}\ket{H_4}\ket{H_5}\ket{H_6}\textrm{.}
\end{equation}
Note that the exact state depends on the particular solution out of
the two possible solutions to the system of equations given in
Eqn.~(\ref{onelost}). Here we are just using Eqn.~(\ref{p1p2}) as a
convenient algebraic way for checking simultaneously both possible
solutions arising from different initial input states.

After the action of the three PBS45$^\circ$ (c.f. dotted line on
Fig.~\ref{circuit} the part of the state that can still contribute
towards a final correct detection pattern is:
\begin{eqnarray}\label{lost1}
&&\frac{1}{8}\left(\ket{H_1} +(p_1-p_2)\ket{V_1}\right)\nonumber
\\
&&~\otimes\left(\ket{H_3H_4} +\ket{V_3V_4}\right)\left(\ket{H_5H_6}
+\ket{V_5V_6}\right)\textrm{.}
\end{eqnarray}

To see why this is so we need to closely examine the action of the
PBS45$^\circ_{12}$ acting on $\ket{H_1}$ and $\ket{H_2}$. This is
summarized below:

\begin{eqnarray}\label{PBS45}
&PBS45^\circ_{12}:\nonumber\\
&\ket{H_1} \rightarrow\frac{1}{2} \left(\ket{H_1}+ \ket{V_1}+ \ket{H_2} - \ket{V_2}\right)\textrm{,} \nonumber \\
&\ket{H_2} \rightarrow \frac{1}{2} \left(\ket{H_1}- \ket{V_1}+
\ket{H_2} + \ket{V_2}\right)\textrm{.}
\end{eqnarray}

Eqns.~(\ref{onelost}) make sure that the state
$p_1\ket{H_1}+p_2\ket{H_2}$ contains at most one photon. Thus by
linearity we can apply the action summarized in Eqn.~(\ref{PBS45})
to deduce that the state after the action of the PBS45$^\circ_{12}$
evolves to:
 \begin{eqnarray}
&&\frac{1}{2} \left[\right. p_1\left(\ket{H_1}+ \ket{V_1}+ \ket{H_2} - \ket{V_2}\right) \nonumber \\
&&~+~p_2\left(\ket{H_1}- \ket{V_1}+ \ket{H_2} + \ket{V_2}\right)\left.\right]\textrm{.}%
\end{eqnarray}
Hence after applying Eqns.~(\ref{onelost}) this becomes:
\begin{equation}
\frac{1}{2} \left(\ket{H_1}+ (p_1-p_2)\ket{V_1}
+\ket{H_2}+(p_2-p_1)\ket{V_2}\right)\textrm{.}
\end{equation}
However the only part of this state that can yield a final correct
detection pattern is:
\begin{equation}
\frac{1}{2} \left(\ket{H_1}+ (p_1-p_2)\ket{V_1}\right)\textrm{.}
\end{equation}

The reason lies in the fact that the 3 pairs of input modes shown in
Fig.~\ref{circuit} must each provide one of the photons which
finally ends up in a detector, in order for a successful detection
outcome to be possible.

To see this consider the example where there is no contribution from
pair 1. This means that the only way that the detector at mode 4
could click with one and only one photon is if only one photon is
present in mode 4 after the action of PBS45$^\circ_{34}$. The reason
for this is that if there are two photons in this mode after the
action of the PBS45$^\circ_{34}$ then necessarily they must be in
the same polarization as indicated by eqn.~(\ref{T2}). That would
suggest that \emph{both} or \emph{none} of them will end up in the
detector in mode 4 yielding an unaccepted detection pattern. If
however only one photon ends up in mode 4 after the action of
PBS45$^\circ_{34}$ (and continues in this spatial mode until it
reaches the detector), suggests that the two remaining photons
initiated in pair 3 (modes 5 and 6, c.f. Fig.~\ref{circuit}) must
somehow end up being detected at the detectors found in modes 1 and
6 (one photon in each detector). However, This is impossible since
if both photons are in mode 6 straight after the action of the
PBS45$^\circ_{16}$ then they will necessarily be in the same
polarization mode according to eqn.~(\ref{T2}). Thus they will both
end up in the same detector - either \emph{both} in mode 1 or
\emph{both} in mode 6. This leads to the conclusion that if a photon
is lost from pair 1 of input modes, then the surviving photon
\emph{must} end up in a detector in order to have any chance of an
overall correct detection pattern in the gate.

It turns out that the same logic can be applied to the cases of
pairs 2 and 3 of input modes, and therefore one can conclude that
the only possible way for receiving a correct detection pattern for
the gate is only if all three photons that are finally detected in
the three separate detectors registering a correct detection outcome
originate from a distinct pair of input modes.

Now that we have established that Eqn.~(\ref{lost1}) gives the state
which can lead to a correct detection outcome after the action of
the three PBS45$^\circ$, and after proceeding with the action of the
PBS$_{14}$, the resulting state is:
\begin{eqnarray}
&&\frac{1}{8}\left[\ket{H_1H_3H_4}
+(p_1-p_2)\ket{V_4H_3H_4}+\ket{H_1V_3V_1}\right.
\nonumber \\
&&~\left.+~(p_1-p_2)\ket{V_1V_3V_4}\right)\left(\ket{H_5H_6}
+\ket{V_5V_6}\right]\textrm{.}
\end{eqnarray}
Post-selecting on the part of the state which has one and only one
photon left in mode 4 gives:
\begin{eqnarray}
&&\frac{1}{8}\left(\ket{H_1H_3H_4}+(p_1-p_2)\ket{V_1V_3V_4}\right)\nonumber\\%
&&~\otimes\left(\ket{H_5H_6}+\ket{V_5V_6}\right)\textrm{.}
\end{eqnarray}
Comparing with Case I, we are at the same stage when the state that
could possibly yield a correct detection pattern was given by
Eqn.~(\ref{beforerotators}). The difference now is that in this case
a photon in mode 2 is missing (plus the factor of $p_1-p_2$ which is
equal to 1 if initially the a photon was absent from mode 2, and -1
if it was absent from mode 1).

By following the same steps as we did for Case I we find that the
term in the state which survives to give a correct detection pattern
corresponding to $\ket{H_1H_4H_6}$ is:
\begin{equation}
\frac{\ket{H_1H_4H_6}}{16\sqrt{2}}\left(\ket{H_3H_5}+(p1-p2)\ket{V_3V_5}
\right)\textrm{.}
\end{equation}
This suggests that the probability with which the initial 5 photon
state ends up with $\ket{\Phi^+}_{35}$ (if initially the photon in
mode 2 was missing) or $\ket{\Phi^-}_{35}$ (if initially the photon
in mode 1 was missing) at the output with all three detectors firing
with one horizontally polarized photon is 1/256. Note that the
remaining possible seven detection outcomes involving one photon
detected at a separate detector occur with the same probability of
1/256 and they give rise to states that are locally equivalent to
$\ket{\Phi^+}_{35}$ or $\ket{\Phi^-}_{35}$ depending which photon
from the initial pair was lost. Thus the overall probability of
ending up with $\ket{\Phi^+}_{35}$ or $\ket{\Phi^-}_{35}$ at the
output is 1/32.

Following the same logic one can show that for the rest of the four
possibilities of operating the gate with 5 photons ( i.e the four
cases where photons in initial modes 3, 4, 5 or 6 are not present)
can lead to an output state $\ket{\Phi^+}_{25}$,
$\ket{\Phi^-}_{25}$, $\ket{\Phi^+}_{23}$ or $\ket{\Phi^-}_{23}$
again with a probability of 1/32. Since the probability of having 5
photons at the input is given by $\eta_S^5(1-\eta_S)$ suggests that
$\ket{2}\bra{2}$ in Eqn.~(\ref{mix}) is given by:
\begin{eqnarray}
\ket{2}\bra{2}&=&\frac{\eta_S^5(1-\eta_S)}{32}[\ket{\Phi^+}_{23}\bra{\Phi^+}_{23}+\ket{\Phi^-}_{23}\bra{\Phi^-}_{23}\nonumber \\%
&&~+\ket{\Phi^+}_{25}\bra{\Phi^+}_{25}+\ket{\Phi^-}_{25}\bra{\Phi^-}_{25}\nonumber \\%
&&~+\ket{\Phi^+}_{35}\bra{\Phi^+}_{35}+\ket{\Phi^-}_{35}\bra{\Phi^-}_{35}]\textrm{,}\nonumber \\%
\end{eqnarray}
which is identical to:
\begin{eqnarray}
\ket{2}\bra{2}&=&\frac{\eta_S^5(1-\eta_S)}{32}[ \ket{H_2H_3}\bra{H_2H_3}+\ket{V_2V_3}\bra{V_2V_3}\nonumber \\
&&~+\ket{H_2H_5}\bra{H_2H_5}+\ket{V_2V_5}\bra{V_2V_5}\nonumber \\
&&~+\ket{H_3H_5}\bra{H_3H_5}+\ket{V_3V_5}\bra{V_3V_5}]\textrm{.}\nonumber \\
\end{eqnarray}
This is just the sum of all possible ways of beginning with a
5-photon input state and ending up with a correct detection pattern.

\subsection*{Case III: 2 photons lost}

We will now look at the case where one photon is lost from pair 1
(modes 1 and 2) and one photon from pair 2 (modes 3 and 4). Remember
that we already eliminated the possibility of having 2 photons lost
from the same pair as with such a scenario is impossible to end up
with one photon arriving at each separate detector which is the
indication of a correct measurement pattern.

Similarly to Case II we will define the variables $p_1$, $p_2$,
$p_3$, $p_4$ for each of the input modes 1, 2, 3 and 4 respectively,
which are described by the equations:
\begin{eqnarray}\label{twolost}
p_1+p_2&=&1\textrm{,} \nonumber \\
p_1p_2&=&0\textrm{,} \nonumber\\
p_3+p_4&=&1\textrm{,} \nonumber \\
p_3p_4&=&0\textrm{.}
\end{eqnarray}
Eqns.~(\ref{twolost}) make sure that two photons are lost in total
and that one photon is lost from each of the first and second pair
of input modes. Note that Eqns.~(\ref{twolost}) have 4 distinct
solutions leading to four distinct 4-qubit states at the input of
the gate given by:
\begin{equation}\label{p1p2p3p4}
(p_1\ket{H_1}+ p_2\ket{H_2})(p_3\ket{H_3} +
p_4\ket{H_4})\ket{H_5}\ket{H_6}\textrm{.}
\end{equation}

Similarly to Case I, Eqn.~(\ref{p1p2p3p4}) is used as a convenient
algebraic way for checking simultaneously the four possible input
states that have one photon missing from each of the first two input
pairs of modes to see with what probability and what state is
received at the output of the gate when a correct detection pattern
is received.

After the action of the three PBS45$^\circ$ (c.f. dotted line on
Fig.~\ref{circuit}) the part of the state that can still contribute
towards a final correct detection pattern is:
\begin{eqnarray}\label{lost2}
&&\frac{1}{8}\left[\left(\ket{H_1}+(p_1-p_2) \ket{V_1}\right)\left(\ket{H_4} +(p_4-p_3)\ket{V_4}\right)\right.\nonumber \\%
&&~\otimes\left.\left(\ket{H_5H_6}+\ket{V_5V_6}\right)\right]\textrm{.}
\end{eqnarray}
This follows from the same logic as for Case I where we showed that
the only way possible that could lead to a correct detection outcome
is if the photon originating from the pair of input modes 1 or 2,
was in mode 1 after the action of the PBS45$^\circ_{12}$.

Furthermore it can be shown that after the action of the
PBS45$^\circ_{34}$ the state in modes 3 and 4 is:
\begin{equation}
\frac{1}{2} \left(\ket{H_3}+ (p_3-p_4)\ket{V_3}
+\ket{H_4}+(p_4-p_3)\ket{V_4}\right)\textrm{.}
\end{equation}
However the only part of this state that can yield a final correct
detection pattern is the one where the surviving photon is in mode 4
given by:
\begin{equation}
\frac{1}{2} \left(\ket{H_4}+ (p_4-p_3)\ket{V_4}\right)\textrm{,}
\end{equation}
since the photon \emph{has to} end in a detector. Thus
Eqn.~(\ref{lost2}) gives the part of the state that may lead to a
correct detection pattern after the action of the three
PBS45$^\circ$.

After the action of the PBS$_{14}$ the state in Eqn.~(\ref{lost2})
evolves to:
\begin{eqnarray}
\frac{1}{8}[(\ket{H_1H_4}+(p_1-p_2)\ket{V_4H_4} +
(p_4-p_3)\ket{H_1V_1} \nonumber \\%
~+(p_1-p_2)(p_4-p_3)\ket{V_1V_4})(\ket{H_5H_6}+\ket{V_5V_6})]\textrm{.}\nonumber\\
\end{eqnarray}
Of this only the parts involving 1 photon in mode 4 can contribute
towards a correct detection outcome. This is given by:
\begin{eqnarray}
&&\frac{1}{8}[(\ket{H_1H_4}+ (p_1-p_2)(p_4-p_3)\ket{V_1V_4})\nonumber \\%
&&~\otimes(\ket{H_5H_6}+\ket{V_5V_6})]\textrm{.}
\end{eqnarray}

After applying PBS$_{16}$ this becomes:
\begin{eqnarray}
\frac{1}{8}[(\ket{H_1H_4H_5H_6}+ (p_1-p_2)(p_4-p_3)\ket{V_6V_4H_5H_6})\nonumber \\%
~+(p_1-p_2)(p_4-p_3)\ket{V_1V_4V_5V_6}+\ket{H_1H_4V_5V_1})]\textrm{.}\nonumber \\
\end{eqnarray}

Post-selecting only the components that can lead to one detection at
each detector in modes 1 and 6 gives:
\begin{equation}
\frac{1}{8}[(\ket{H_1H_4H_5H_6}+
(p_1-p_2)(p_4-p_3)\ket{V_1V_4V_5V_6}]\textrm{.}
\end{equation}

By applying the three 45$^\circ$-polarization rotators found prior
to the three detectors at modes 1, 4 and 6 we find that the only
term that survives involving $\ket{H_1H_4H_6}$ is:
\begin{equation}
\frac{\ket{H_1H_4H_6}}{16\sqrt{2}}\left(\ket{H_5}+(p_1-p_2)(p_4-p_3)\ket{V_5}
\right)\textrm{.}
\end{equation}

This suggests that the probability with which the initial 4-photon
state ends up with the state $\ket{+_5}$ at the output with all
three detectors firing with one horizontally polarized photon is
$2\left[\frac{1}{16}\right]^2=1/128$ (since there are two possible
ways for $(p_1-p_2)(p_4-p_3)=1$. Note that
$\ket{+_i}=1$/$\sqrt{2}\left(\ket{H_i} \pm \ket{V_i}\right)$. By
considering the fact that there are 8 different successful detection
outcomes then the probability of obtaining the $\ket{+_5}$ state is
1/16 since each of the successful outcomes happens with probability
1/128. Also note that 1/16 is also the probability for getting a
correct detection pattern corresponding to the output state
$\ket{-_5}$. This happens for input states where
$(p_1-p_2)(p_4-p_3)=-1$ holds.

This entire procedure can be repeated to demonstrate that when a
photon is lost from the pair of input modes 3 and 4, and one photon
from the pair of input modes 5 and 6 then with probability 1/16 we
end up with state $\ket{+_2}$ at the output or with probability 1/16
we end up with state $\ket{-_2}$ at the output depending on the
combination that the 2 photons are lost. Similarly we can show that
with probability 1/16 we can have either the state $\ket{+_3}$ or
$\ket{-_3}$ at the output depending on which photon is lost from
each of the pairs of input modes 1 and 2 and input modes 5 and 6.

Since the probability of having 4 photons at the input is given by
$\eta_S^4(1-\eta_S)^2$ suggests that $\ket{1}\bra{1}$ in
Eqn.~(\ref{mix}) is given by:
\begin{eqnarray}
\ket{1}\bra{1}=
\frac{\eta_S^4(1-\eta_S)^2}{16}[\ket{+_2}\bra{+_2}+\ket{-_2}\bra{-_2}~~~~~~~{}\nonumber\\%
+\ket{+_3}\bra{+_3}+\ket{-_3}\bra{-_3}+\ket{+_5}\bra{+_5}+\ket{-_5}\bra{-_5}]\textrm{.}
\end{eqnarray}
which is identical to:
\begin{eqnarray}
\ket{1}\bra{1}=
\frac{\eta_S^4(1-\eta_S)^2}{16}[\ket{H_2}\bra{H_2}+\ket{V_2}\bra{V_2}~~~~~~~{}\nonumber\\%
~+\ket{H_3}\bra{H_3}+\ket{V_3}\bra{V_3}+\ket{H_5}\bra{H_5}+\ket{V_5}\bra{V_5}]\textrm{.}
\end{eqnarray}
This is the sum up all possibilities of beginning with a 4-photon
input state and ending up with a correct detection pattern.

\subsection*{Case IV: 3 photons lost}

We will now look at the case where one photon is lost from each of
the three pairs of input modes in the gate shown in
Fig.~\ref{circuit}. In this case we will define the six variables
$p_1$...$p_6$ which can be described by the equations:
\begin{eqnarray}\label{threelost}
p_1+p_2&=&1\textrm{,} \nonumber \\
p_1p_2&=&0\textrm{,} \nonumber\\
p_3+p_4&=&1\textrm{,} \nonumber \\
p_3p_4&=&0\textrm{,} \nonumber\\
p_5+p_6&=&1\textrm{,} \nonumber \\
p_5p_6&=&0\textrm.{}
\end{eqnarray}
The solution to the system of equations given in
Eqns.~(\ref{threelost}), make sure that three photons are lost in
total and that one and only one photon is lost from each of the
three pairs of input modes. There are 8 distinct solutions which
lead to 8 distinct 3-photon input states at the gate which can be
collectively and conveniently described by:
\begin{eqnarray}
&&(p_1\ket{H_1}
+p_2\ket{H_2})(p_3\ket{H_3}+p_4\ket{H_4})\nonumber\\%
&&~~~~\otimes(p_5\ket{H_5}+p_6\ket{H_6})\textrm{.}
\end{eqnarray}

After the action of the three PBS45$^\circ$ (c.f. dotted line on
Fig.~\ref{circuit}) and following exactly the same logic as we did
for the previous cases, the part of the state that can still
contribute towards a final correct detection pattern is:
\begin{eqnarray}
&&\frac{1}{8}\left[\left(\ket{H_1}+(p_1-p_2) \ket{V_1}\right)\left(\ket{H_4} +(p_4-p_3)\ket{V_4}\right)\right.\nonumber \\
&&~\left.\left(\ket{H_6}+p_6-p_5){V_6}\right)\right]\textrm{.}
\end{eqnarray}

After the action of the PBS$_{14}$ followed by the PBS$_{16}$ the
part of the state that can lead to three separate detector clicks is
given by:
\begin{equation}\label{lost3}
\frac{1}{8}[(\ket{H_1H_4H_6}+
(p_1-p_2)(p_4-p_3)(p_6-p_5)\ket{V_1V_4V_6}]\textrm{.}
\end{equation}

By applying the 3 45$^\circ$-polarization rotators found prior to
the three detectors at modes 1, 4 and 6 we find that the only term
that survives involving $\ket{H_1H_4H_6}$ is:
\begin{equation}
\frac{1+(p_1-p_2)(p_4-p_3)(p_6-p_5)}{16\sqrt{2}}\ket{H_1H_4H_6}\textrm{.}
\end{equation}

This term has a non-zero amplitude only if:
\begin{equation}\label{check}
1+(p_1-p_2)(p_4-p_3)(p_6-p_5)\neq 0\textrm{.}
\end{equation}
There are four solutions to Eqns.~(\ref{threelost}) that satisfy
Eqn.~(\ref{check}). These are the solutions satisfying:
\begin{equation}
(p_1-p_2)(p_4-p_3)(p_6-p_5)=1\textrm{,}
\end{equation}
suggesting that the probability by which one horizontally polarized
photon can be registered at each separate detector is
$4\left[\frac{2}{16\sqrt{2}}\right]^2=1/32$. As there are 8
different accepted detection outcomes all of which can be shown to
happen with a probability of 1/32 as well then the overall success
probability for receiving a correct measurement pattern is 1/4.

Since the probability of having 3 photons at the input is given by
$\eta_S^3(1-\eta_S)^3$ suggests that $\ket{0}\bra{0}$ in
Eqn.~(\ref{mix}) is given by:
\begin{eqnarray}
\ket{0}\bra{0}= \frac{\eta_S^3(1-\eta_S)^3}{4}\ket{vac}\bra{vac}
\textrm{,}
\end{eqnarray}
where $\ket{vac}$ refers to the vacuum state.
\newline

\subsection*{Determining $\rho_{out}$}

The un-normalized density operator for the output state is given by:
\begin{equation}
\rho_{out}=\sum_{i=0}^3\ket{i}\bra{i}\textrm{,}
\end{equation}

This can be derived from sum of all the possibilities for receiving
a correct measurement pattern at the detectors of the gate shown in
Fig.~\ref{circuit} given that we have perfect detectors and single
photon sources of efficiency $\eta_S<1$. In each of the previous
sections we calculated the parts, $\ket{i}\bra{i}$'s, contributing
to the final mixed state where a different but fixed number of
photons was lost at the input of the gate for each of those cases
(Cases I to IV). After summation and re-normalization the resulting
output mixed state can be shown to be:

\begin{eqnarray}
\rho_{out}&=& \frac{\eta_S^3}{(2-\eta_S)^3} \ket{GHZ_{235}}\bra{GHZ_{235}}\nonumber \\
&+&\frac{\eta_S^2(1-\eta_S)}{(2-\eta_S)^3}(\ket{H_2H_3}\bra{H_2H_3}+\ket{V_2V_3}\bra{V_2V_3}\nonumber\\%
&&~~~~~~~~~~~~~~~~~~~\ldots+\ket{V_3V_5}\bra{V_3V_5}) \nonumber\\
&+&\frac{2\eta_S(1-\eta_S)^2}{(2-\eta_S)^3}\left(\ket{H_2}\bra{H_2}+\ket{V_2}\bra{V_2}\right. \nonumber\\%
&&~~~~~~~~~~~~~~~~~~~\left.\ldots+\ket{V_5}\bra{V_5}\right) \nonumber\\
&+&\frac{8(1-\eta_S)^3}{(2-\eta_S)^3}\ket{vac}\bra{vac}\textrm{.}
\end{eqnarray}

Comparing with Eqn.~(\ref{ghz}) we find that $\rho_{out}$ is
identical to a GHZ-ID, where $f=\frac{\eta_S}{2-\eta_S}$ which
proves that the proposed gate shown in Fig.~\ref{circuit} is capable
for producing this independently degradable state for the case where
perfect detectors and imperfect sources are used.

\subsection*{Equivalence of different scenarios}

In this section we will demonstrate that the scenario where we have
perfect detectors and imperfect sources of efficiency $\eta_S'$ is
exactly equivalent to the scenario where we have both imperfect
sources and detectors with respective efficiencies $\eta_S$ and
$\eta_D$ such that $\eta_S' =\eta_S\eta_D$.

\begin{figure}[h]
\begin{center}\includegraphics[width=8.2cm]{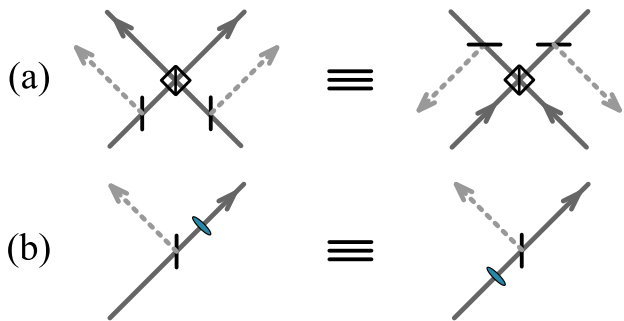}\end{center}
\caption{ \label{commutations} (a) Losses occurring at the two input
spatial modes of a beamsplitter can be commuted to the output if the
loss rate is the same on both modes (and vice versa). (b) Losses
occurring at the output of a polarization rotator can be commuted to
the input. In both (a) and (b) losses are modeled by variable
beamsplitters reflecting part of the signal in non-computational
modes indicated by dotted lines.}
\end{figure}

By considering the commutation relations represented by
Fig.~\ref{commutations} one can show that losses occurring at the
three detectors shown in Fig.~\ref{circuit} can be commuted all the
way to the sources. To show this we will first assume that the three
detectors share the same efficiency $\eta_D$. We will also simulate
each inefficient detector as equivalent to a perfect detector with
the addition of variable beamsplitter (VARBMS) placed directly at
the input of the detector which has transmissivity $\eta_D$. In
other words with probability $1-\eta_D$ the VARBMS reflects an
incoming signal into a non computational mode. The task is to show
that these three VARBMS's can be commuted all the way to the
sources. Clearly they can be commuted all the way to the output of
the three PBS45$^\circ$.

To see this consider first the two VARBMS's at the input of the
detectors in modes 1 and 6. These can be commuted all the way to the
input of the PBS$_{16}$ by applying the commutation relations given
by Fig.~\ref{commutations}. By then there would be two VARBMS's of
the same reflectivity at the output of PBS${14}$, thus applying the
commutation relations again can lead to the case where we have one
VARBMS at modes 1,4 and 6 right at the output of the three
PBS$45^\circ$'s (c.f. dotted line in Fig.~\ref{circuit}).

It is important to appreciate now that ultimately the output at
modes 2,3 and 5 will lead to a detector. If these have the same
efficiency as the detectors in the gate shown in Fig.~\ref{circuit}
suggests that the VARBMS's modeling their efficiency can commuted
all the way to the outputs of the PBS$45^\circ$'s shown in
Fig.~\ref{circuit}. This in fact holds true even if there is a bit
of linear optics circuitry lying in between the detectors that are
ultimately at the end of modes 2, 3 and 5. This is because it is
possible to commute the VARBMS's right to the output of the three
PBS$45^\circ$'s of the gate shown in Fig.~\ref{circuit} since one
can commute the VARBMS through any intermediate linear optics
circuits involving just PBS's, polarization rotators, or phase
shifters.

After this stage there would be one VARBMS at each output of the the
three PBS$45^\circ$'s that share the same tnasmissivity, $\eta_D$
(assuming of course that the detector where modes 2,3 and 5 end at
have the same efficiency $\eta_D$ as those at the end of modes 1,4
and 6 do. Therefore the 6 VARBMS's would be commuted all the way to
the sources.

If each of the source has efficiency $\eta_S$ then this again can be
modeled by a perfect source with a VARBMS of reflectivity $\eta_S$
placed directly on the source's output mode which partly reflects
the photons in some non computational spatial mode. In this case we
will have two VARBMS's aligned which can be modeled as a single
VARBMS with transmissivity $\eta_S\eta_D$. This can now actually be
use to simulate the scenario where we have perfect detectors and
imperfect sources which have a source efficiency $\eta_S'
=\eta_S\eta_D$ and thus this proves the equivalence of the two
scenarios.

\subsection*{Conclusion}

We have seen therefore that the circuit of Fig.~\ref{circuit} is
capable of turning an imperfect source of 6 photons with a loss rate
$\eta_S$, and detectors of finite efficiency $\eta_D$ into an
ID-GHZ. The ``equivalent effective loss rate" of the ID-GHS state
is:
\begin{equation}
\frac{\eta_S\eta_D}{2-\eta_S\eta_D}\textrm{.}
\end{equation}

\end{appendix}


\begin{thebibliography}{99}
\bibitem{barrettweaknon} Although recent work (see e.g. \cite{barrett}) has lessened this requirement to a more managable but still challenging figure.
\end{thebibliography}
\end{document}